\documentclass[a4paper,11pt]{article}
\usepackage{pos}

\title{Profound optical flares from the relativistic jets of active galactic nuclei}
 \ShortTitle{blazar flares}

\author[a]{Gopal Bhatta}
\author[b]{Staszek Zola}
\author[c]{M. Drozdz}
\author[d]{Daniel Reichart}
\author[d]{Joshua Haislip}
\author[d]{Vladimir Kouprianov}
\author[e]{Katsura Matsumoto}
\author[f,g]{Eda Sonbas}
\author[h]{D. Caton}
\author[b]{Urszula Pajdosz-\'Smierciak}
\author[i]{A. Simon}
\author[j,k]{J. Provencal}
\author*[a]{Dariusz G\'ora}
\author[b]{Grzegorz Stachowski}

\affiliation[a]{Institute of Nuclear Physics Polish Academy of Sciences, 
  PL-31342 Krak\'ow, Poland}

\affiliation[b]{Astronomical Observatory of the Jagiellonian University,
ul. Orla 171, 30-244 Krak\'ow, Poland}

\affiliation[c]{Mt. Suhora Observatory, Pedagogical University,
ul. Podchorazych 2, 30-084 Krak\'ow, Poland}

\affiliation[d]{Dept. of Physics and Astronomy, University of North Carolina at Chapel Hill,
Chapel Hill, NC 27599, USA}

\affiliation[e]{Astronomical Institute, Osaka Kyoiku University,
4-698 Asahigaoka, Kashiwara, Osaka 582-8582, Japan}

\affiliation[f]{University of Adiyaman, Department of Physics,
02040 Adiyaman, Turkey}

\affiliation[g]{Astrophysics Application and Research Center, Adiyaman University, Adiyaman 02040, Turkey}

\affiliation[h]{Dark Sky Observatory, Dept. of Physics and Astronomy, Appalachian State University,
 Boone, NC 28608, USA}
 
 \affiliation[i]{Astronomy and Space Physics Department, Taras Shevshenko National University of Kyiv,
 Volodymyrska str. 60, 01033 Kyiv, Ukraine}
 
  \affiliation[j]{University of Delaware, Department of Physics and Astronomy Newark,
 DE 19716, USA}
 
  \affiliation[k]{Delaware Asteroseismic Research Center, Mt. Cuba Observatory,
 Greenville, DE 19807, USA}

\emailAdd{dariusz.gora@ifj.edu.pl}
\abstract{Intense outbursts in blazars are among the most extreme phenomena seen in extragalactic objects. Studying these events can offer important information about the energetic physical processes taking place within the innermost regions of blazars, which are beyond the resolution of current instruments. This work presents some of the largest and most rapid flares detected in the optical band from the sources 3C 279, OJ 49, S4 0954+658, Ton 599, and PG 1553+113, which are mostly TeV blazars. The source flux increased by nearly ten times within a few weeks, indicating the violent nature of these events. Such energetic events might originate from magnetohydrodynamical instabilities near the base of the jets, triggered by processes modulated by the magnetic field of the accretion disc. We explain the emergence of flares owing to the injection of high-energy particles by the shock wave passing along the relativistic jets. Alternatively, the flares may have also arisen due to geometrical effects related to the jets. We discuss both source-intrinsic and source-extrinsic scenarios as possible explanations for the observed large amplitude flux changes.}

\ConferenceLogo{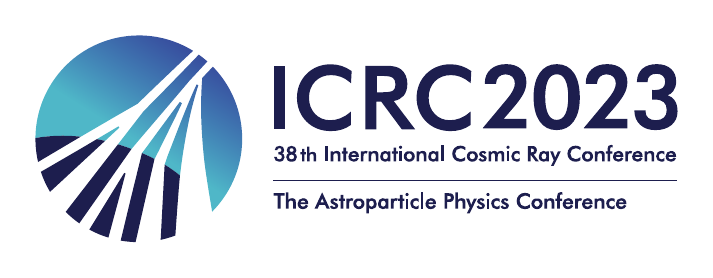}

\FullConference{%
38th International Cosmic Ray Conference (ICRC2023)\\
  26 July - 3 August, 2023\\
  Nagoya, Japan}

\begin{document}
\maketitle

\section{Introduction}
Blazars, a sub-class of radio-loud active galactic nuclei (AGN), exhibit intense outbursts and extreme phenomena in extragalactic objects. They feature relativistic jets closely aligned with the line of sight, which serve as sources of highly variable non-thermal continuum emission that is Doppler boosted \citep{Urry1995}. Blazars can be divided into two sub-classes: flat-spectrum radio quasars (FSRQ), which exhibit broad emission lines, and BL Lacertae (BL Lac) objects, which show weak or no emission lines. Despite their differences, both types are visible in the TeV energy range and are the predominant discrete gamma-ray sources in the sky.  The broadband non-thermal spectrum of blazars extends from radio to TeV gamma-rays and consists of low- and high-energy components. The low-energy component arises from synchrotron emission by relativistic plasma in the magnetized jets. Various models, including leptonic and hadronic scenarios, have been proposed to explain the origin of the high-energy emission. Leptonic models involve ultra-relativistic electrons up-scattering low-energy photons via the inverse Compton mechanism. In the synchrotron self-Compton model (SSC; e.g. \citep{Marscher1985,Maraschi1992}), synchrotron photons produced by electrons are inverse-Compton scattered by co-spatial leptons. External Compton (EC) models suggest contributions from AGN components for inverse-Compton scattering. on the other hand, hadronic models involve high-energy protons producing the high-energy spectral component through proton-synchrotron and/or photon-initiated cascades  \citep[see][]{Ghisellini1996,Sikora1994,Dermer1992}. While both models can explain the emission, hadronic models require a stronger magnetic field to cool more massive protons \citep{Mucke2003,Aharonian2000,Mannheim1993}. Additionally, proton-synchrotron models require the acceleration of ultra-high-energy cosmic rays (UHECRs) in blazar jets, resulting in gamma-ray emission and neutrino production. 

Multiple wavelength (MWL) observations reveal that blazars emit  flux at  that is variable across a wide range of timescales, e. g. from decades to a few minutes \citep[ see][]{Bhatta2018a,bhatta16b, Bhatta2018c, Bhatta2020, Bhatta2021}. While the general statistical nature of variablity can be approximated by a single power-law of spectral density  (see optical and $\gamma$-ray \citep[][]{Nilsson2018,Bhatta2020}, blazars often display complex variability patterns, such as red-noise-like variability combined with quasi-periodic oscillations (QPO) and occasional sharp flux rises. These distinct flaring events, lasting from weeks to months, indicate extreme physical conditions in the central engine and jets, driving efficient particle acceleration and cooling processes. Gamma-ray flares in blazars are frequently accompanied by rotation of optical polarization angle  and ejection of superluminal radio knots \citep{Blinov2018}. Flaring events are associated with disturbances propagating along the jet, which lead to particle energization through shock waves in weakly magnetized jets, or through magnetic reconnection and shock wave re-collimation in highly magnetized jets. \citep{ Jorstad2016, Lind1985, Giannios2013, Bromberg2009}.

In this proceeding, we present a modeling approach for analyzing the flaring observations of five blazars, which were obtained through long-term optical monitoring of AGN conducted by our research group. In Section \ref{sec:2}, we provide a detailed account of the observations of the source sample and the relevant data processing procedures for the optical data. Then, in Section \ref{sec:5}, we discuss the details of the models used to explain the flares observed in the  light curves. The outcomes of the results and potential implications  are summarized in Section \ref{sec:4}.

\begin{table}[b]
 \caption{General properties the sample of blazar targets \label{table:1} }
 \centering
 \begin{tabular}{l|l|l|l|c}
 \hline
 Source name & Source class &R.A. (J2000) & Dec. (J2000) & Redshift (z) \\
 \hline
 OJ 49 &BL Lac, LSP & $08^h31^m48.88^s$ & $+04^d29^m39.086^s$ & 0.17386\\
		S4 0954+658 & BL Lac, LSP& $09^h58^m47.2^s$ & $+65^d33^m55^s$ & 0.368 \\ 
 TXS 1156+295 &FSRQ, LSP & $11^h59^m032.07^s$ & $+29^d14^m42.0^s$ & 0.729 \\
 3C 279 &FSRQ, LSP& $12^h56^m11.1665^s$ & $-05^d47^m21.523^s$ & 0.536 \\
 PG 1553+113 &BL Lac, HSP& $15^h55^m43.044^s$ & $+11^d11^m24.365^s$ & 0.36 \\ 
 \hline
 \end{tabular}
 \end{table}
 
 \section{Observations and data processing }
 \label{sec:2}
To constrain the variability properties of AGN, several telescopes have been employed to monitor a sample of quasars over an extended duration through the Skynet Robotic Telescope Network \citep{Skynet}. Optical band observations of the sources 3C 279, OJ 49, S4 0954+658, Ton 599, and PG 1553+113 were acquired through the network. Furthermore, additional data were collected from telescopes in Turkey, Japan, and Poland. The study primarily used the wide band R filter (Bessell prescription), with longer runs conducted at the Krakow and Mt. Suhora sites.

The data from the Skynet consisted of scientific images taken each night, which were then processed by the network pipeline to remove bias, dark, and flatfield effects. Raw images from other sites were accompanied by calibration frames. The standard procedure of calibration, including bias, dark, and flatfield correction using the IRAF package, was applied to the raw images. Magnitude extraction was performed using aperture photometry with the CMunipack program, which employs the DAOPHOT algorithm. Differential magnitudes were derived by comparing the objects with selected comparison stars visible in the field of view of all telescopes. Moreover, the constancy of the comparison stars was verified using check stars chosen in a similar manner.

Table \ref{table:1} presents the sample of sources with their classes, positions, and redshifts, along with the blazar source classification based on the synchrotron peak frequency. Table \ref{table:2} lists the total observation duration, number of observations, and mean magnitude for each source.
 
 \begin{table*}
 \caption{Optical observations of a sample of blazars and their variability properties}
 \centering
 \label{table:2}
 \begin{tabular}{l|r|l|l|c|c|c|c|c}
 \hline
 Source name & Duration (d) &Npt.& mean mag& VA (mag)&Fvar (\%)&t$_{\rm var}$ (min.)\\
 (1)& (2) &(3)& (4)& (5)&(6)&(7)\\
 \hline
 OJ 49 &164.28& 620& 16.28 & 2.75&66.23$\pm$0.24&38.24$\pm$11.80\\
 TXS 1156+295 &79.65 & 650 & 17.45 & 3.30&62.78$\pm$0.13 &19.06$\pm$13.73\\
 PG 1553+113 &97.43 & 1071& 15.57& 0.91 & 23.61$\pm$0.21&64.76$\pm$11.18\\
 3C 279 &461.58&1831 & 17.59 & 2.78&46.16$\pm$0.50 &11.73$\pm$7.80\\
 S4 0954+658 &242.70& 2988 & 14.80 & 2.61 &47.60$\pm$0.17&17.10$\pm$6.18\\ 
 \hline
 \end{tabular}
\end{table*}

\begin{figure*}
\begin{center} 
{\includegraphics[width=0.48\textwidth,angle=0]{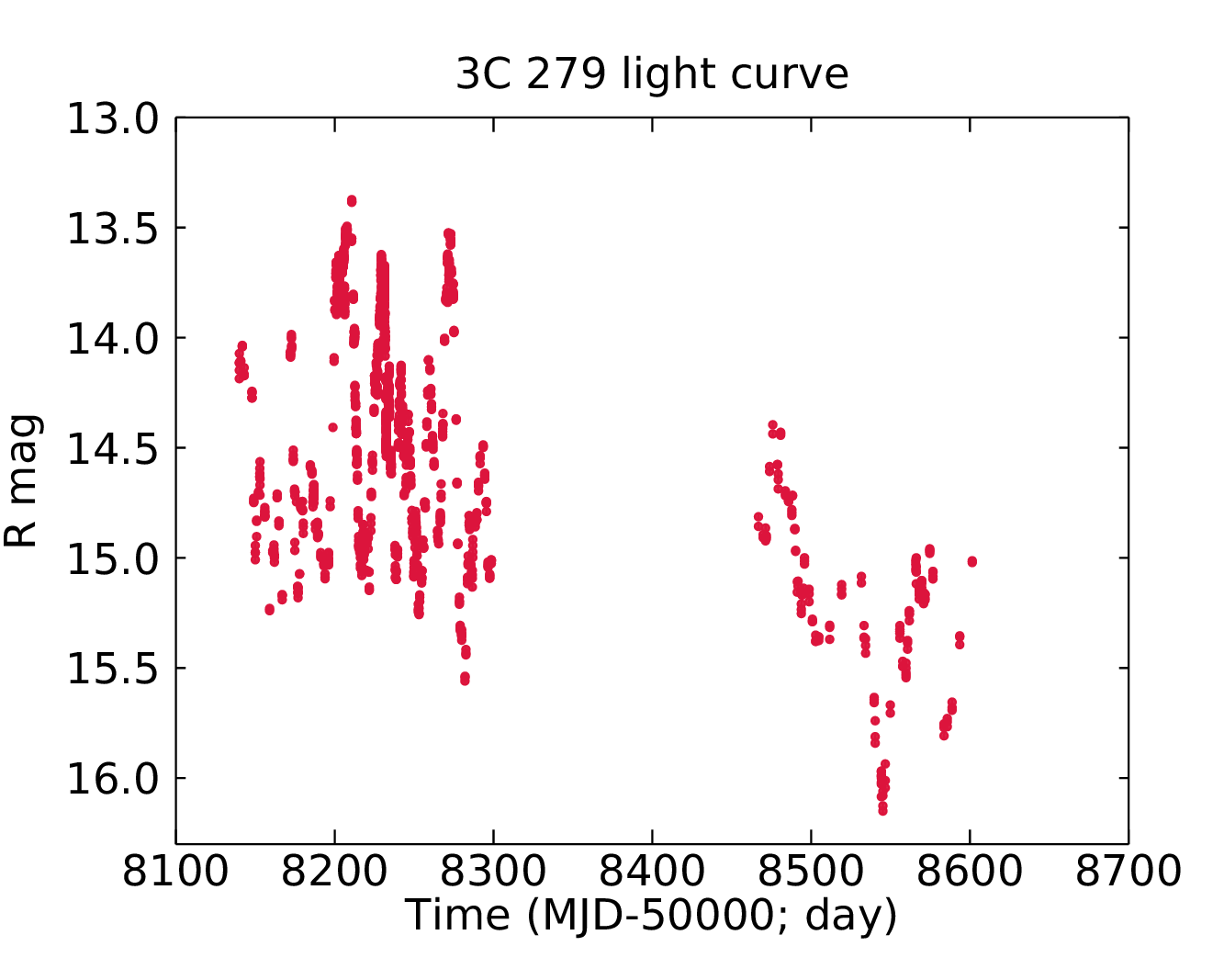}} 
\hspace{-0.1cm}
{\includegraphics[width=0.48\textwidth,angle=0]{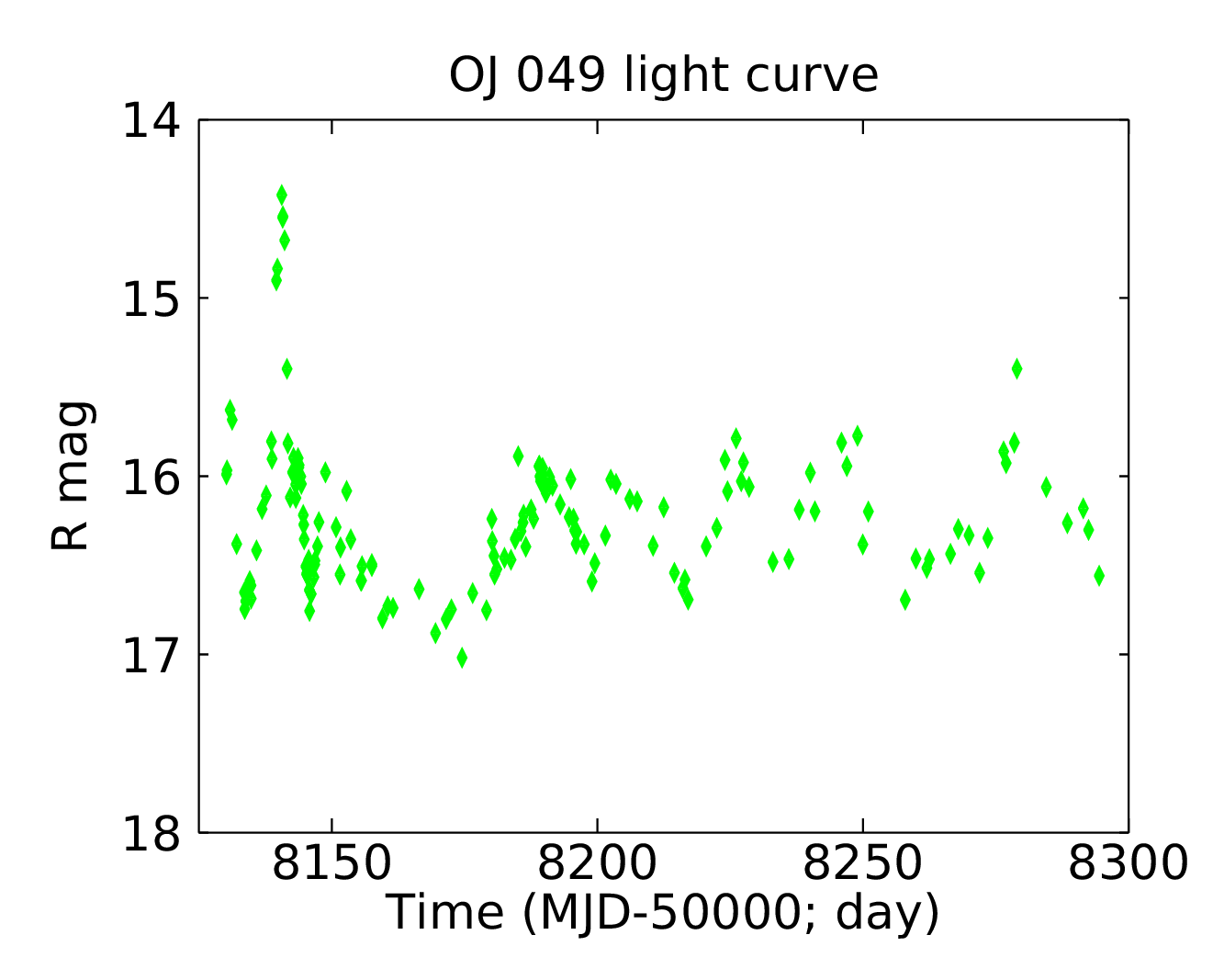}}\\
{\includegraphics[width=0.48\textwidth,angle=0]{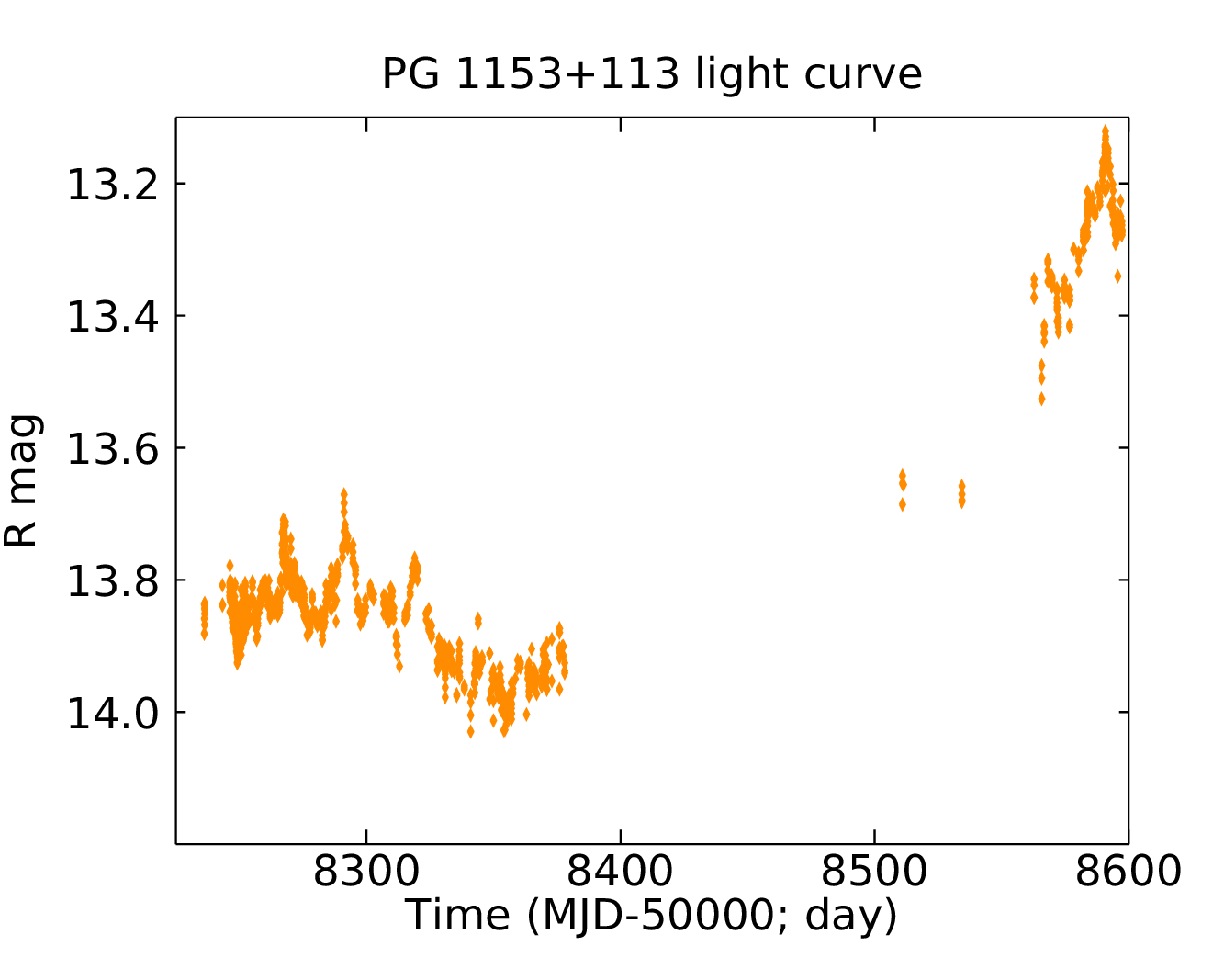}}
\hspace{-0.1cm}
{\includegraphics[width=0.48\textwidth,angle=0]{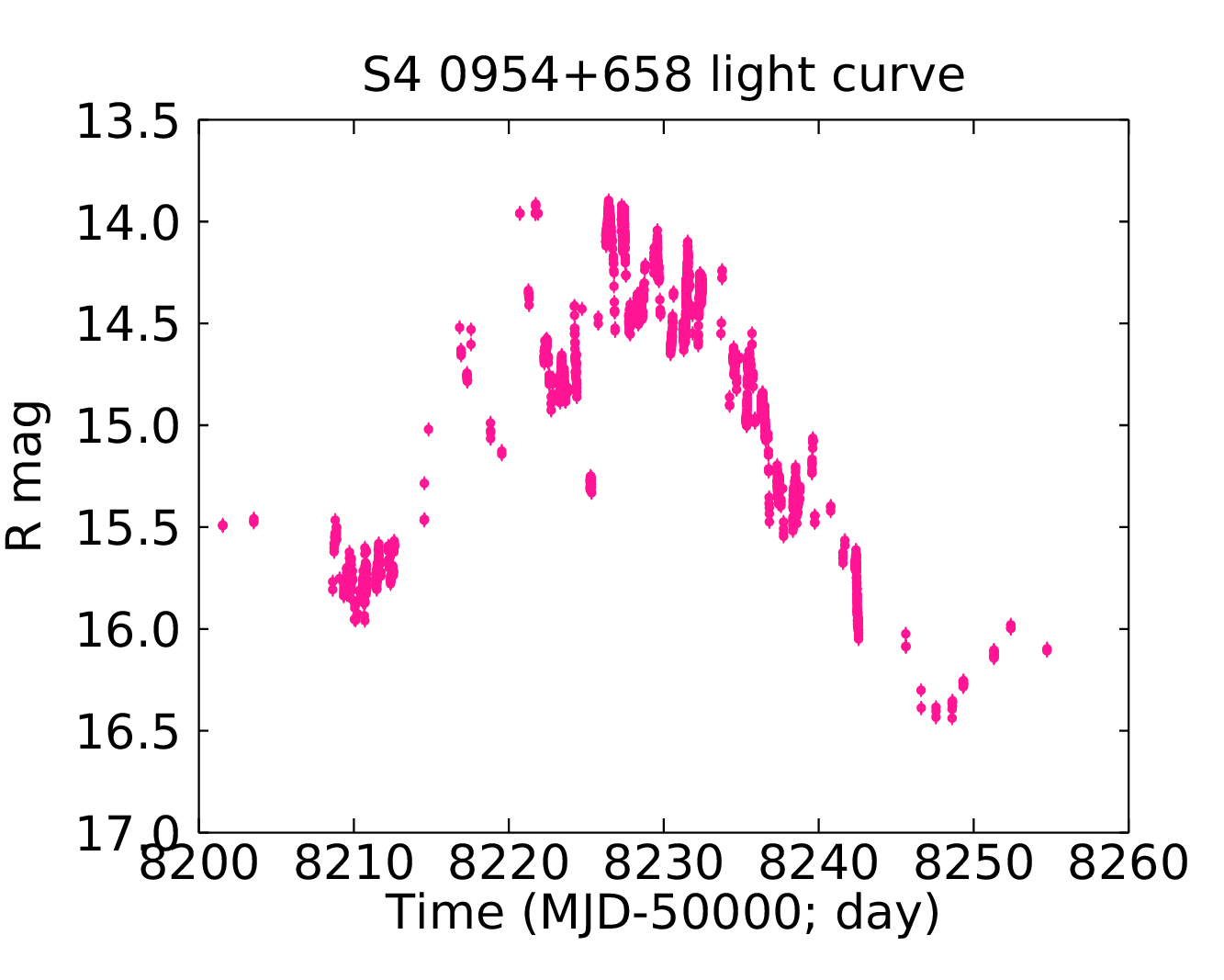}}\\
{\includegraphics[width=0.48\textwidth,angle=0]{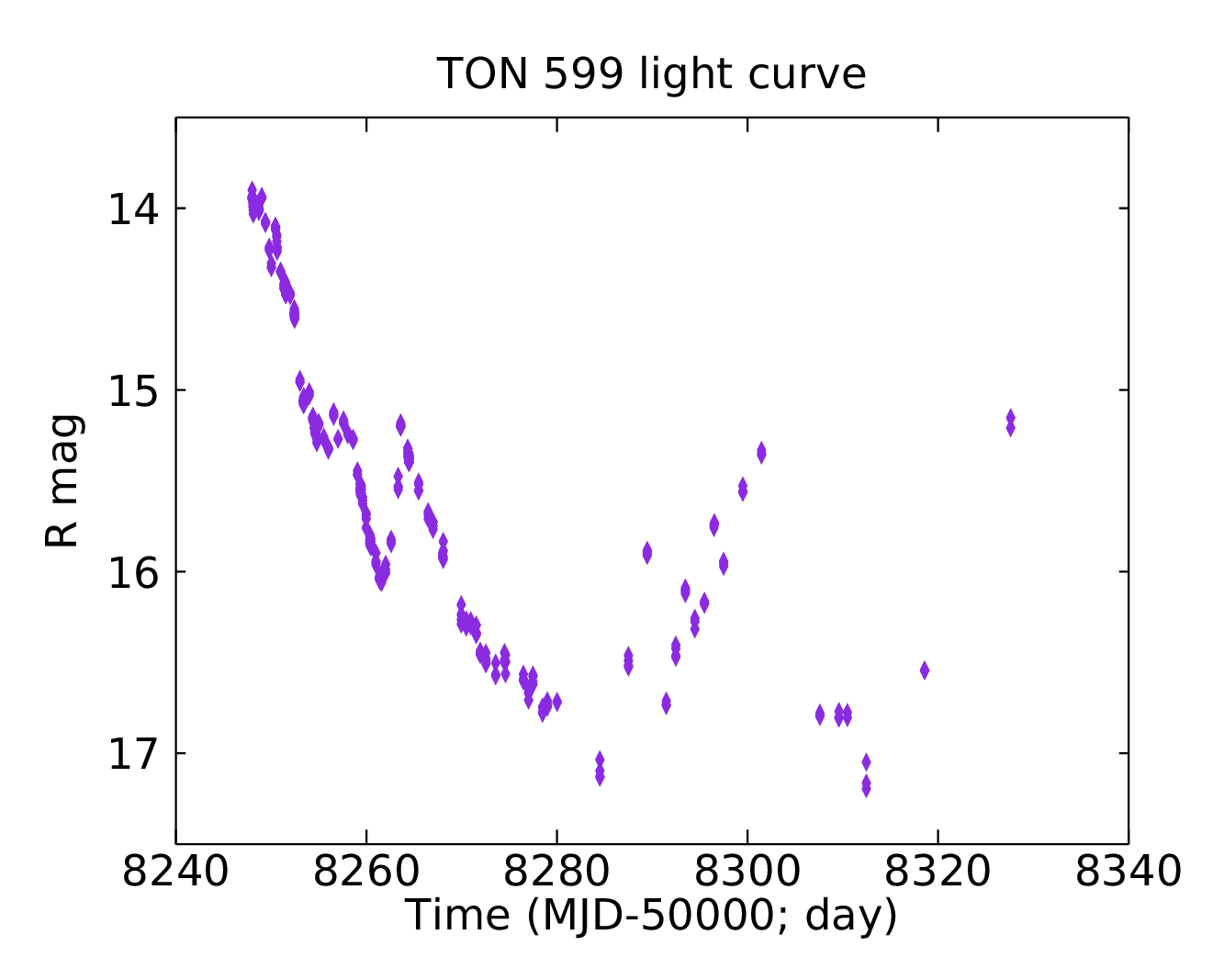}}
\hspace{-0.1cm}
{\includegraphics[width=0.48\textwidth,angle=0]{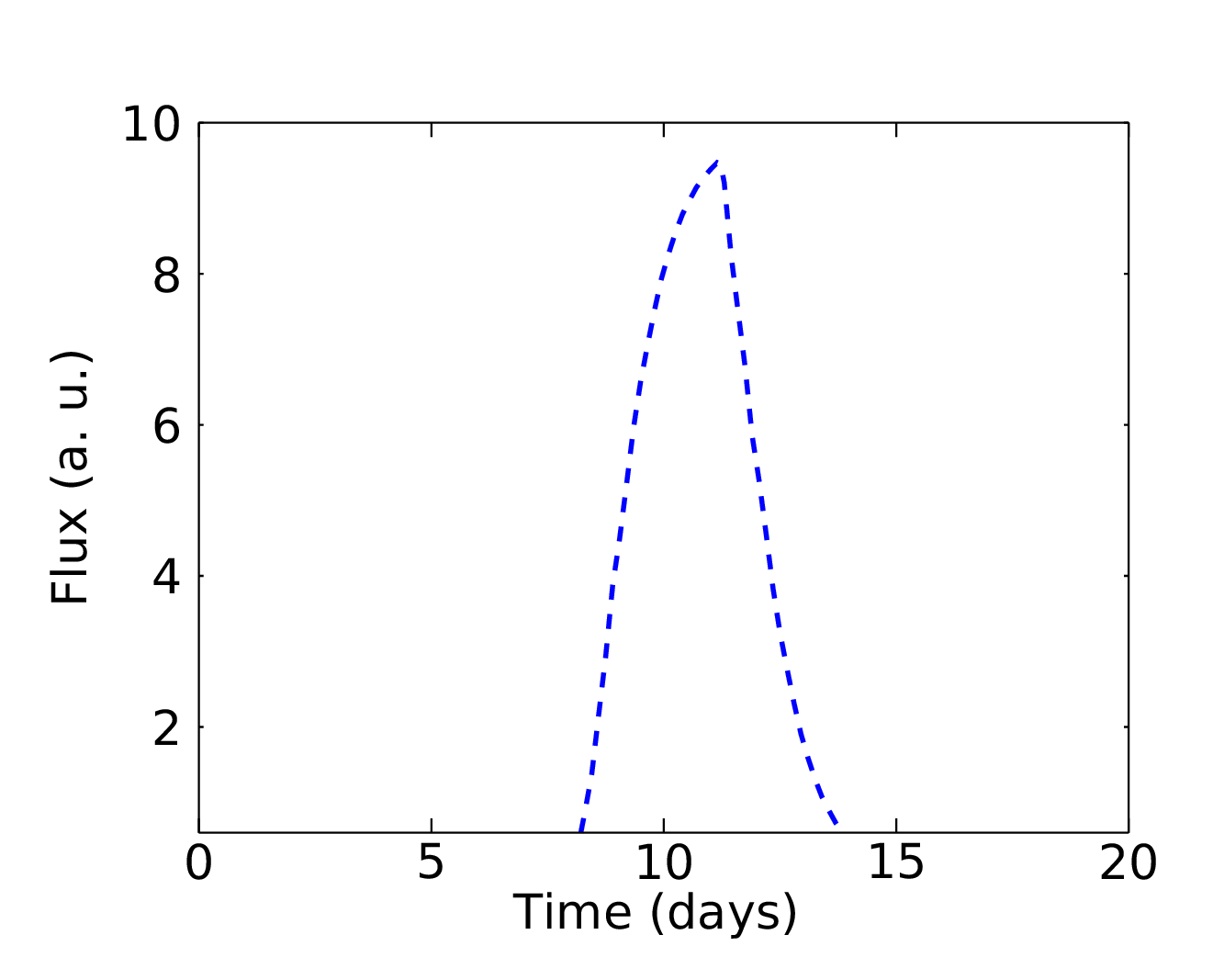}}
\caption{R-band optical observations of the sample of blazars. The bottom right plot shows the emergence of flares in blazars via shock-in-jet processes. }
\label{fig:1}
\end{center}
\end{figure*}

\section{ Analysis and Modeling}
\label{sec:5}
The light curves presented in Figure 1 clearly demonstrate significant flux variability in the sources. This variability, quantified using fractional variability \cite{Vaughan2003}, is listed in the 6th column of Table \ref{table:2}. Similarly, a measure of the variability timescale, represented by the minimum variability time scale \cite{Bhatta2018c}, is listed in the 7th column of the table. Furthermore, the modeling of the flares using both intrinsic and extrinsic source scenarios is presented in the following sections.
\subsection{ Source intrinsic scenario}
Blazar flares can be explained within the scenario of internal shocks moving along their relativistic jets. To demonstrate this, we attempt to create a flare lasting about a week using the approach described \citet[][]{Kirk1998} (for details see \cite{Bhatta2023}). The method involves a time-dependent analysis of a homogeneous single zone leptonic model, where variable emission results in significant flares due to gradual particle acceleration at the shock front and subsequent radiative cooling at the emission region. The model assumes cylindrical jets aligned near the line of sight, with particles being injected at a constant rate at the shock wavefront and primarily losing energy through synchrotron emission.

 The evolution of particles in the acceleration zone can be described by the following form of the diffusion equation 
\begin{equation}
\frac{\partial N}{\partial t}+\frac{\partial }{\partial \gamma }\left [ \left ( \frac{\gamma }{t_{acc}}-\beta _{s} \gamma ^{2} \right ) \right ]N +\frac{N}{t_{acc}}=Q\delta (\gamma -\gamma _{0}),
\label{diffusion1}
\end{equation}
where $\beta _{s} \gamma ^{2}$ represents the loss of the energy by the synchrotron radiation given by
\begin{equation}
\beta _{s}=\frac{4}{3}\frac{\sigma_{ T}}{m_{e}c^{2}}\left ( \frac{B^{2}}{2\mu _{0}} \right ).
\label{energy}
\end{equation}
Here $\sigma_{T}$ represents the Thompson-scattering cross-section, B denotes the magnetic field, and $\mu_{0}$ stands for the permeability of free space. Furthermore, the following equations are used to represent the particle enhancement in the regions,
%\begin{equation}
$ Q\left ( t \right )=Q_{0}$ for t $<$ 0 and t $>$ $t_{f}$ and $Q\left ( t \right )=\left ( 1+\eta _{f} \right )Q_{0}$ for 0 $<$ t $<$ $t_{f}$. Then, the corresponding rise in the intensity of the source flux can be given by
\begin{equation}
\label{flare}
I\left ( \nu ,t \right )=I_{1}\left ( \nu ,t \right )+\eta _{f}\left [ I_{1}\left ( \nu ,t \right )-I_{1}\left ( \nu ,\left ( 1-u_{s}/c \right )t_{f} \right ) \right ].
\end{equation}
For small angles, where cos$\theta \rm \sim1$, and with $\delta \sim 2\Gamma$, the relations $I(\nu,t)=\delta^3I(\nu',t') \sim 8\Gamma^3I(\nu',t')$, and $\nu =\Gamma \left ( 1+\beta \right )\approx 2\Gamma \nu' $ are utilized to convert the source rest-frame quantities (primed) to the observer's frame (unprimed).
In this context, we consider $\nu=4.55\times10^{14}$ Hz, which corresponds to the mean effective wavelength of R-band (658 nm), and the shocks travel down the blazar jet at relativistic speeds ($\beta_{s}=0.1$). The resultant normalized intensity profile, which imitates the flaring behavior observed in the blazars, is displayed in the bottom right panel of Figure \ref{fig:1}

\begin{figure*}
\centering
\includegraphics[width=0.47\textwidth,angle=0]{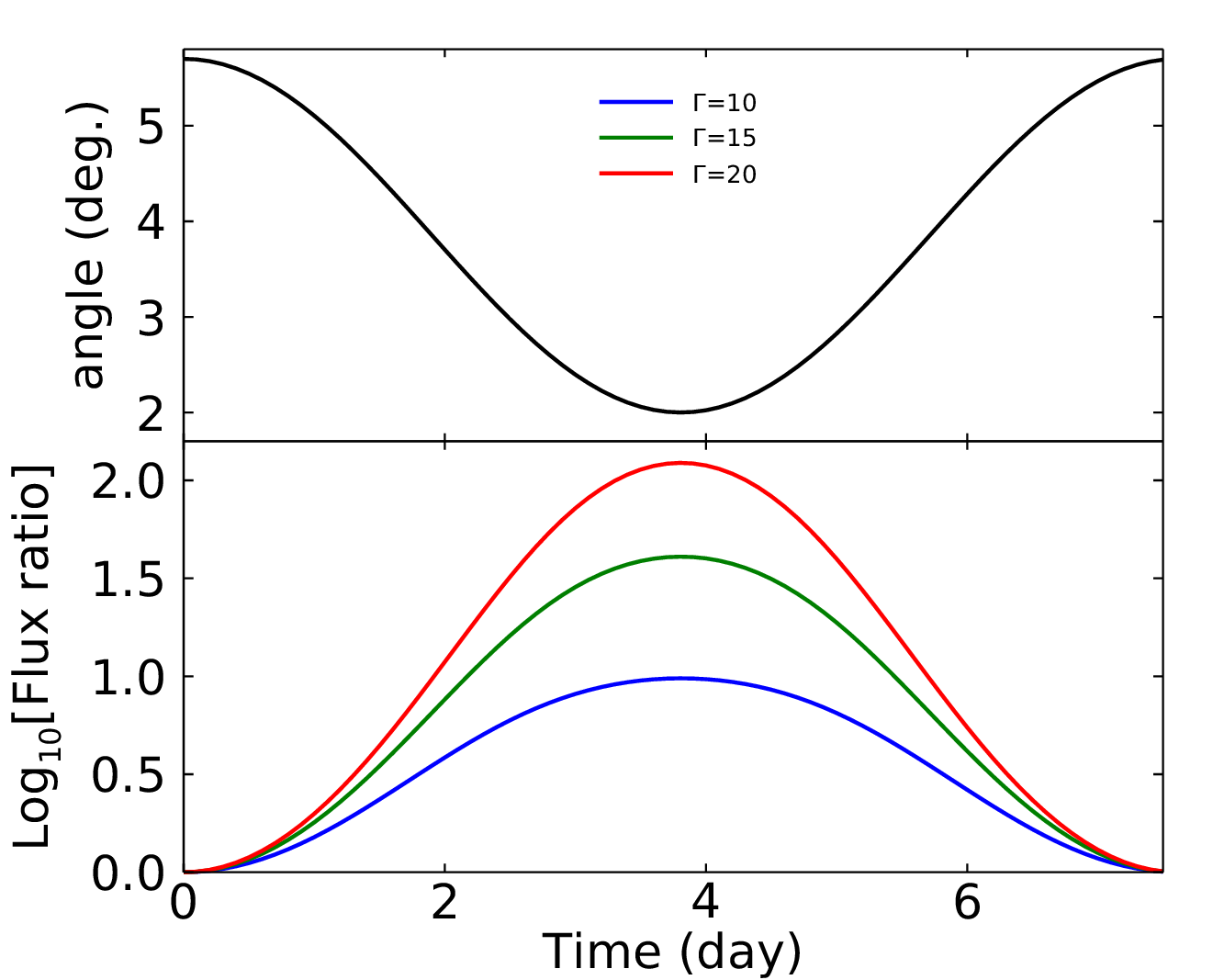}
\includegraphics[width=0.47\textwidth,angle=0]{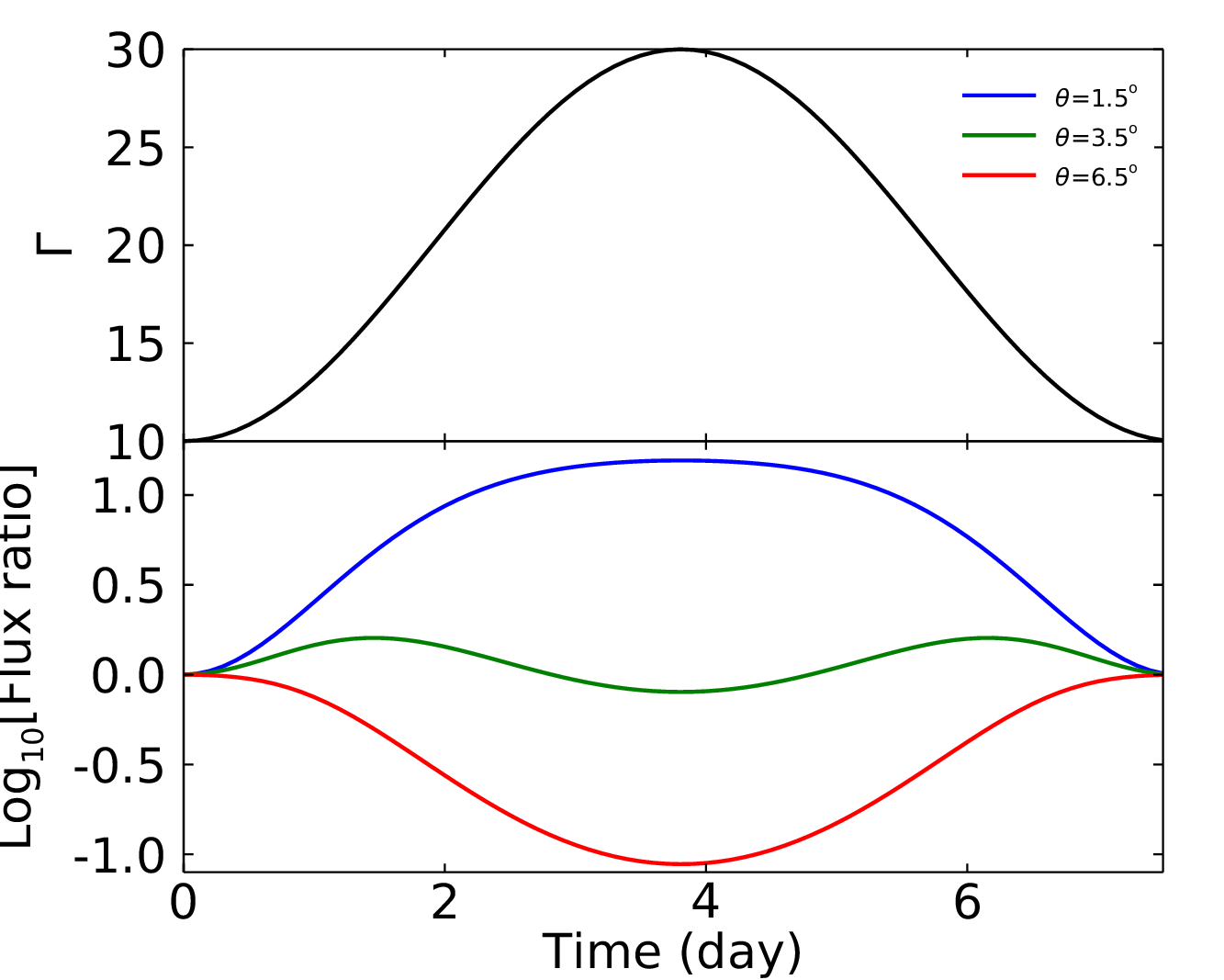}
 \caption{\textit{Left:} As the angle between the emission region and the line of sight decreases (top panel), the flux appears to flare as a result of relativistic beaming (bottom panel). The three curves correspond to the three different values of the bulk Lorentz factors. \textit{Right:} Similarly, flaring of flux (bottom panel) owing to the increase in the bulk Lorentz factor (top panel). The three curves correspond to the three different values of the angles of sight.}
\label{fig:beam}
\end{figure*}

\subsection{ Source extrinsic scenario}

Large flares in blazars can also be attributed to the Doppler-boosted emission resulting from emission regions moving along curved trajectories within the jets. In such scenarios, the flux in the rest frame (${F}'{{\nu'}}$) is connected to the observed flux ($F{\nu}$) using the following equations:

 \begin{equation}
\frac{F_{\nu}(\nu)}{{F}'_{{\nu}'} (\nu)}=\delta^{3+\alpha} \quad\text{and}\quad \delta(t)=\frac{1}{\Gamma \left ( 1-\beta cos\theta \right )}
\label{flux}
\end{equation}

In general, the optical spectral slope tends to be greater than 1 in Low-Synchrotron-Peaked (LSP) blazars, such as OJ 49, S4 0954+658, TON 599, and 3C 279, while it is less than 1 in High-Synchrotron-Peaked (HSP) blazars, specifically in this case, PG1553+113, serves as an example with such behavior. For the purpose of illustration, we assume a spectral index of approximately $\alpha \sim 1$.

Assuming that the apparent flux rise occurs purely due to changes in the Doppler factor ($\delta$), this can be related to alterations in the angle with the line of sight and/or changes in the bulk Lorentz factor ($\Gamma$). Here, for illustrative purposes, we consider two cases: one involving changes in $\theta$ and another involving changes in $\Gamma$.
\begin{itemize}

\item Change in the angle of the line of sight:
To simulate a gradual decrease in the angle between the emission region and the line of sight, we use the approximation $\rm{ \theta=\theta_{0}-Asin^{2}\omega t}$, where $ \theta_{0}$ is set to 5.7 and A is set to 3.7, as depicted in the top left panel of Figure \ref{fig:beam}. The resulting flux rise profiles for three values of the bulk Lorentz factor, namely $\Gamma$=10, 15, and 20, are shown as blue, green, and red curves, respectively, in the lower left panel of Figure \ref{fig:beam}. 

\item Change in $\Gamma$:
In this scenario, the bulk Lorentz factor of the dominant emission region gradually increases following the approximation $\rm{ \Gamma=\Gamma_{0}+Asin^{2}\omega t}$, with an initial value of $ \Gamma_{0}$=10 and A=20. This allows the plasma blob traveling at a speed of $ \Gamma=10$ to accelerate and reach $ \Gamma=30$, then decelerate back to its original speed. The evolution of $ \Gamma$ over time is presented in the top right panel of Figure \ref{fig:beam}.  The resulting flux rise profiles for three values of the angles of the line of sight, i.e. $\theta$=1.5, 3.5, and 6.5$^o$, are depicted as blue, green, and red curves, respectively, in the lower right panel of Figure \ref{fig:beam}.
\end{itemize}

\section{ Discussion \& Conclusion}
\label{sec:4}
Blazars exhibit violent variability across a wide range of timescales, from minutes to decades. Their light curves often display well-defined MWL  flaring events, characterized by a sudden increases in blazar flux followed by energetic dissipation, which usually lasts for a few weeks to months. Numerous blazar monitoring studies focus on gamma-ray flares frequently linked to the ejection of radio knots visible in VLBA images \citep{Agudo2011,Jorstad2013}. Emerging evidence suggests that gamma-ray flares may be associated with superluminally-moving features passing through stationary features along the jet; many flares coincide with the passage of superluminal knots through the millimeter-wave core \citep[see][and the references therein]{Jorstad2017}.  Particle acceleration mechanisms, including shock diffusive acceleration and turbulent jet models \citep{Blandford1987,Hughes98,Marscher14,Narayan2012}, play a crucial role in accelerating particles to millions of Lorentz factors. In the turbulent jet scenario, shock waves in the relativistic plasma's turbulent flow can heat and compress particles, causing individual turbulent cells to appear as flares. Similarly, high magnetic fields in the jet can trigger intermittent magnetic instabilities, leading to magnetic reconnection events and particle acceleration (\citep{Spruit2001,Giannios2019}. These events can produce rapid flares with distinctive envelopes, possibly showing slight asymmetry due to disturbance passing through the emission region or geometric effects like light crossing time. 

\section*{Acknowledgments}
 We acknowledge the support of the Polish National Science Centre through the grants 2018/29/B/ST9/01793 (SZ), and 2020/39 / B / ST9 / 01398 (DG). KM acknowledges JSPS KAKENHI grant number 19K03930.

%% Full authors list (ONLY FOR COLLABORATIONS)
%\clearpage
%\section*{Full Authors List: \Coll\ Collaboration}
%
%\noindent \textbf{Note comment afterwards:} Collaborations have the possibility to provide an authors list in xml format which will be used while generating the DOI entries making the full authors list searchable in databases like Inspire HEP. \\
%
%\scriptsize
%\noindent
%first.author$^1$, 
%second.author$^2$, 
%third.author$^3$ % .... more names
%and 
%last.author$^{n}$ \\
%
%\noindent
%$^1$first.affiliation.
%$^2$second.affiliation. % .... more affiliation
%$^{m}$last.affiliation.

\end{document}